\documentclass[preprint2]{aastex}
\shorttitle{NGC 1333 IRAS 4A2 Ammonia Disk Kinematics}
\shortauthors{Choi}

\usepackage{times}
\slugcomment{To appear in the Astrophysical Journal}

\begin{document}

\fontsize{10}{10.6}\selectfont

\title{Kinematics of the Ammonia Disk around the Protostar NGC 1333 IRAS 4A2}
\author{\sc Minho Choi\altaffilmark{1}, Ken'ichi Tatematsu\altaffilmark{2},
            and Miju Kang\altaffilmark{1}}
\affil{$^1$ International Center for Astrophysics,
            Korea Astronomy and Space Science Institute,
            Daedeokdaero 776, Yuseong, Daejeon 305-348, South Korea;
            minho@kasi.re.kr.\\
       $^2$ National Astronomical Observatory of Japan,
            2-21-1 Osawa, Mitaka, Tokyo 181-8588, Japan}
\setcounter{footnote}{2}

\begin{abstract}

\fontsize{10}{10.6}\selectfont

The NGC 1333 IRAS 4A protobinary was observed
in the ammonia (2, 2) and (3, 3) lines
with an angular resolution of 0.3 arcsec.
The ammonia emission source of IRAS 4A2 is elongated
in the direction perpendicular to the bipolar jet
and has a size of 0.55 arcsec or 130 AU.
This emission structure was interpreted
as a circumstellar disk around the IRAS 4A2 protostar,
and the rotation kinematics of the disk
was investigated by making a position-velocity diagram along the major axis.
Assuming a Keplerian rotation,
the disk has a rotation velocity of 1.8 km s$^{-1}$ at a radius of 20 AU,
which implies a central object of about 0.08 solar masses.
The collapse age of the protostar is about 50,000 yr.
The mass, accretion rate, and age are consistent with what are expected
from the standard theory of low-mass star formation.
If IRAS 4A2 grows at this rate, it may become a star similar to the Sun.
\end{abstract}


\section{INTRODUCTION}

The NGC 1333 region is a nearby star-forming site
at a distance of 235 pc from the Sun
and contains many young stellar objects
(Aspin et al. 1994; Bally et al. 1996; Rodr\'{\i}guez et al. 1999;
Sandell \& Knee 2001; Hirota et al. 2008).
One of the youngest objects among them is the binary protostar IRAS 4A,
and the luminosity suggests that the protostars are growing vigorously
through the accretion of gas from the molecular cloud surrounding them
(Sandell et al. 1991; Lay et al. 1995; Looney et al. 2000).
The spectral energy distribution suggests
that they are Class 0 protostars
(Sandell et al. 1991; Enoch et al. 2009).

The northwestern component of the binary, IRAS 4A2,
drives a prominent bipolar jet
and is associated with H$_2$O maser sources
(Blake et al. 1995; Lefloch et al. 1998; Choi 2005; Park \& Choi 2007).
The accretion disk of IRAS 4A2 is bright in NH$_3$ lines
and elongated in the direction perpendicular to the jet axis
(Choi et al. 2007).
Since the jet axis is close to the plane of the sky (Choi et al. 2006),
the disk is seen nearly edge-on.
While the blueshifted/redshifted emission peaks of the NH$_3$ maps
are displaced in a way suggestive of rotation (Choi et al. 2007),
the spatial resolution was not high enough to produce a rotation curve
that can be used to derive interesting physical quantities such as mass.

Mass is one of the most fundamental quantities of a stellar object,
and direct estimations of mass
from the rotation kinematics of circumstellar disk
are important in understanding the evolution of young stellar objects.
This is especially true for very young protostars,
but reliable measurements of rotation kinematics
are difficult for many reasons
such as confusion with the protostellar envelope.
For the IRAS 4A protostars,
the NH$_3$ lines are bright and seem to selectively trace the disks
(Choi et al. 2007),
which allow a detailed study.
Other evolutionary indicators,
such as bolometric temperature ($T_{\rm bol}$),
are usually used instead of mass,
and it is important to verify them by measuring mass.

In this Letter, we present the results of
our observations of the NGC 1333 IRAS 4 region
in the NH$_3$ (2, 2) and (3, 3) lines
with an angular resolution higher than that of the previous study
(Choi et al. 2007).
We describe our observations in Section 2.
In Section 3, we report the results of the NH$_3$ imaging.
In Section 4, we discuss the star-forming activities in the IRAS 4A2 region.

\section{OBSERVATIONS}

The NGC 1333 IRAS 4 region was observed using the Very Large Array (VLA)
of the National Radio Astronomy Observatory
in the NH$_3$ (2, 2) and (3, 3) lines
(23.7226336 and 23.8701296 GHz, respectively).
Twenty-seven antennas were used
in the B-array configuration.
Three tracks of observations were performed in 2008 January.
For each of the NH$_3$ lines,
the spectral windows were set to have 64 channels
with a channel width of 0.049 MHz,
giving a velocity resolution of 0.62 km~s$^{-1}$.

The phase tracking center was ($\alpha$, $\delta$)
= (03$^{\mathrm h}$29$^{\mathrm m}$10\fs413,
31\arcdeg13$'$32\farcs20) in J2000.0.
The nearby quasar 0336+323 (PKS 0333+321) was observed
to determine the phase and to obtain the bandpass response.
The flux was calibrated by observing the quasar 0713+438 (QSO B0710+439)
and by setting its flux density to 0.55 Jy
(VLA Calibrator Flux Density Database%
\footnote{See http://aips2.nrao.edu/vla/calflux.html.}).
Comparison of the amplitude gave a flux density of 0.95 Jy for 0336+323,
and the flux uncertainty is $\sim$10\%.
To avoid the degradation of sensitivity owing to pointing errors,
pointing was referenced
by observing the calibrators at the $X$ band ($\lambda$ = 3.6 cm).
This referenced pointing was performed
about once an hour and just before observing the flux calibrator.

Maps were made using a CLEAN algorithm.
The images were made with the B-array data only.
With a natural weighting,
the NH$_3$ data produced a synthesized beam
of 0\farcs33 $\times$ 0\farcs28 in full-width at half-maximum (FWHM),
but the deconvolution maps were restored with a circular beam of 0\farcs30
for the ease of analysis.

\section{RESULTS}

\begin{figure*}[!t]
\epsscale{2}
\plotone{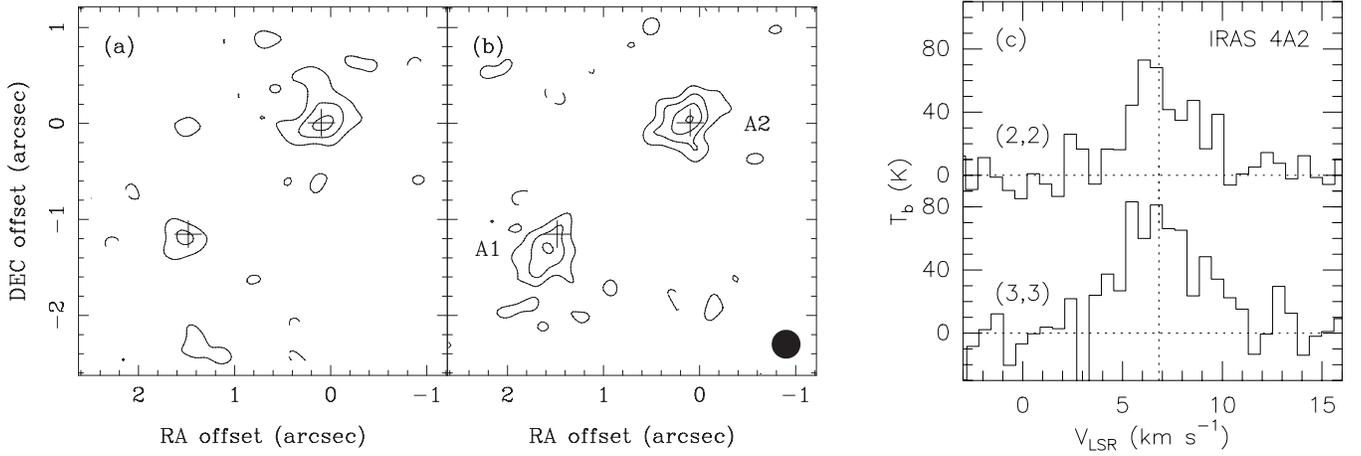}
\caption{\small\baselineskip=0.825\baselineskip
Maps and spectra of the NH$_3$ lines toward the NGC 1333 IRAS 4A region.
(a)
Map of the NH$_3$ (2, 2) line.
The line intensity was averaged
over the velocity interval of $V_{\rm LSR}$ = (4.6, 8.8) km s$^{-1}$.
The contour levels are 1, 2, 3, and 4 $\times$ 0.6 mJy beam$^{-1}$,
and the rms noise is 0.2 mJy beam$^{-1}$.
Dashed contours are for negative levels.
(b)
Map of the NH$_3$ (3, 3) line.
Map parameters are the same as (a).
Shown in the bottom right-hand corner is the restoring beam:
FWHM = 0\farcs3.
Plus signs:
the 3.6 cm continuum sources (Reipurth et al. 2002).
(c)
Spectra of the NH$_3$ lines toward IRAS 4A2,
at the peak positions of the 3.6 cm continuum emission.
Contribution from the dust emission was subtracted
by determining the continuum flux level
in the velocity intervals of (--6.8, --0.7) and (14.1, 20.2) km s$^{-1}$.
Vertical dotted line:
systemic velocity of IRAS 4A2 ($V_{\rm LSR}$ = 6.83 km~s$^{-1}$)
determined by the model fitting.}
\end{figure*}

\begin{figure}[!t]
\epsscale{0.9}
\plotone{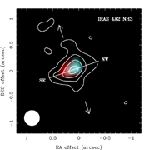}
\centerline{\scriptsize [See http://minho.kasi.re.kr/Publications.html
for the original high-quality figure.]}
\vspace{-0.5\baselineskip}
\caption{\small\baselineskip=0.825\baselineskip
Map of the circumstellar disk of IRAS 4A2.
The NH$_3$ (2, 2) and (3, 3) line maps were averaged.
Contours show the intensity distribution
averaged over the velocity interval of $V_{\rm LSR}$ = (4.6, 8.8) km~s$^{-1}$.
Contour levels are 0.5, 1.0, 1.5, and 2.0 mJy beam$^{-1}$,
and the rms noise is 0.16 mJy beam$^{-1}$.
Cyan and red images show the emission
in the blueshifted (4.6--5.8 km s$^{-1}$)
and redshifted (7.6--8.8 km s$^{-1}$)
parts of the velocity interval, respectively.
Shown in the bottom left-hand corner is the restoring beam:
FWHM = 0\farcs3.
Straight line:
cut for position-velocity diagram (Figure 3).
The cut goes through the peak position and has P.A. = 108\fdg9,
which is perpendicular to the jet axis.
Arrows:
direction of the bipolar jet (Choi 2005).}
\end{figure}

The new NH$_3$ images (Figure 1) have an angular resolution
three times better than the images of the previous study
(Choi et al. 2007).
The two NH$_3$ lines show similar structures and spectra.
Since our analysis does not depend on the specifics of the lines,
images of the two lines were averaged,
which improves the signal-to-noise ratio.
The peak position of IRAS 4A2 in the resulting image (Figure 2),
$\alpha$ = 03:29:10.42 and $\delta$ = 31:13:32.2 (J2000),
coincides with the position of the centimeter continuum peak
(Reipurth et al. 2002).
The deconvolved sizes of the sources from elliptical Gaussian fits
are FWHM = 0\farcs44 $\times$ 0\farcs25
with position angle (P.A.) = 146\arcdeg\ for IRAS 4A1
and FWHM = 0\farcs55 $\times$ 0\farcs30 with P.A. = 129\arcdeg\ for IRAS 4A2.
While the relation between the IRAS 4A1 source and the associated outflow
is not clear,
the IRAS 4A2 source is elongated in the direction
almost exactly perpendicular to the jet axis.
Therefore, we will concentrate on IRAS 4A2 for the rest of this Letter.

The IRAS 4A2 source is clearly resolved along the major axis.
The northwestern side of the disk is blueshifted,
and the southeastern side is redshifted relative to the systemic velocity
(Figure 2).
The position-velocity (PV) diagram
along the major axis of the disk (Figure 3(a))
shows a diagonal ridge of emission
with low-level structures extended along the velocity and position axes,
which is typical of rotation velocity profiles decreasing with radius. 
Since the IRAS 4A2 source has a morphology and kinematics
consistent with a Keplerian-like disk,
most of the emission seems to be coming
from the accretion disk around the protostar.

\begin{figure}[!t]
\epsscale{1}
\plotone{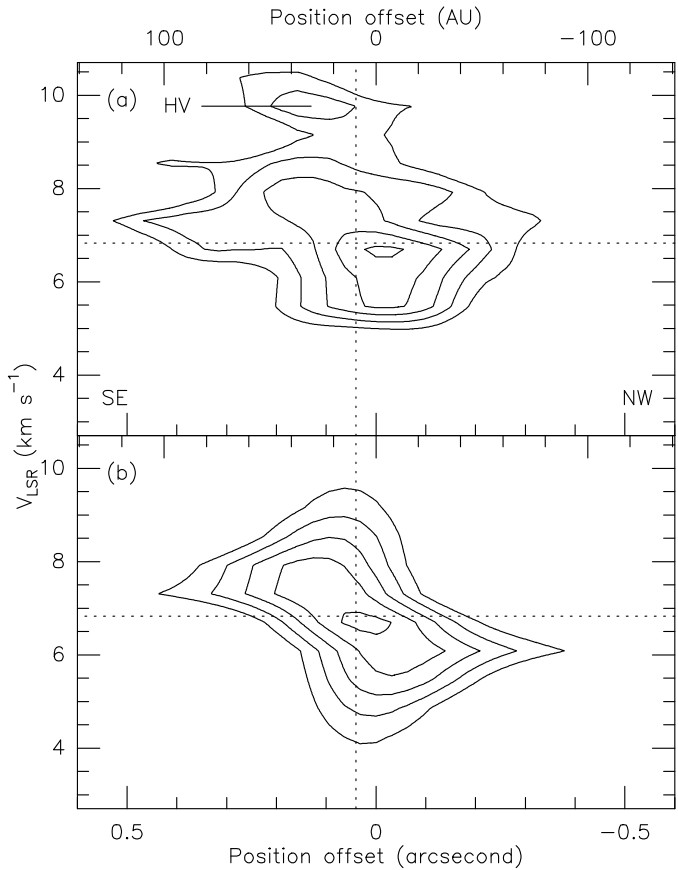}
\caption{\small\baselineskip=0.825\baselineskip
PV diagrams.
(a)
PV diagram from the average NH$_3$ map (Figure 2)
along the major axis of the IRAS 4A2 circumstellar disk.
Contour levels are 3, 4, 5, 6, and 7 times the rms noise (10.3 K).
The high-velocity (HV) component at 9.8 km s$^{-1}$
does not seem to follow the disk kinematics
and was ignored in the model-fitting process.
(b)
PV diagram of the best-fit model of a gas disk with Keplerian rotation.
Dotted lines:
the central position and systemic velocity of the disk
found by the model fit.}
\end{figure}

\begin{figure}[!b]
\epsscale{1.0}
\plotone{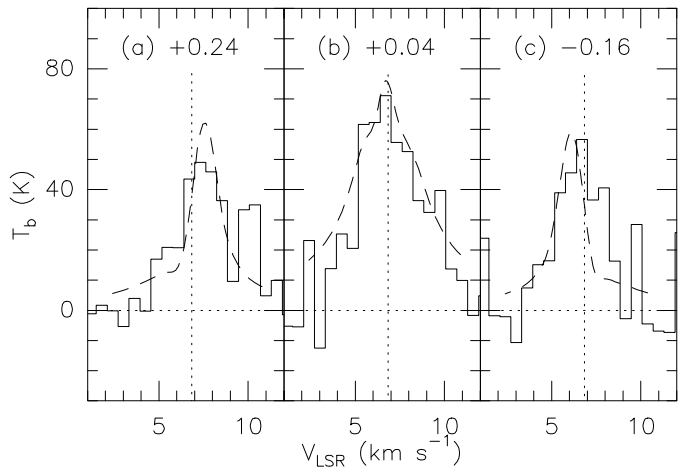}
\caption{\small\baselineskip=0.825\baselineskip
Spectra of the NH$_3$ line, (2, 2) and (3, 3) averaged,
toward three positions on the PV cut.
(a)
Spectrum toward 0\farcs2 southeast of the central position.
(b)
Spectrum toward the central position found by the model fit.
(c)
Spectrum toward 0\farcs2 northwest of the central position.
Dashed spectra:
spectra from the best-fit model.
Vertical dotted line:
systemic velocity ($V_{\rm LSR}$ = 6.83 km s$^{-1}$).}
\end{figure}

\section{DISCUSSION}

\subsection{Disk Model}

Models of a rotating gas disk were constructed
to analyze the PV diagram of the IRAS 4A2 disk (Figure 3(a)).
The binary companion, IRAS 4A1, is at least 430 AU away
and would not affect the kinematics of the IRAS 4A2 disk,
especially the inner part where most of the emission is coming from.
The gas motion was assumed to have
only the azimuthal component with a Keplerian rotation profile,
\begin{equation}
v_\phi(r) \propto r^{-1/2},
\end{equation}
where $v_\phi$ is the rotation velocity, and $r$ is the radius.

Since our main interest is the determination of $v_\phi$,
we tried to make the other aspects of the model as simple as possible.
(For a more sophisticated treatment, see Dutrey et al. 2007.)
The disk was assumed to be geometrically thin and cylindrically symmetric.
The rotation axis of the disk was assumed be the same as the jet axis,
and the line-of-sight velocity of the SiO emission line (Choi 2005)
and the proper motion of the molecular hydrogen outflow (Choi et al. 2006)
give an inclination angle of 10\fdg7.
The central position and systemic velocity
were allowed to vary to fit the PV diagram.

The region of the disk traced by the NH$_3$ lines has inner and outer edges
at radii of $R_{\rm in}$ and $R_{\rm out}$, respectively.
The surface brightness of the NH$_3$ line
was assumed to have a power-law profile,
\begin{equation}
S \propto r^{-a}.
\end{equation}
It was found that the fit
is not very sensitive to $R_{\rm in}$ or $R_{\rm out}$
because $R_{\rm in}$ is much smaller than the beam size
and because the intensity in the outer region slowly fades
below the detection limit.
Any $R_{\rm in}$ less than 0.4 AU is acceptable.
The best-fit $R_{\rm out}$ is 310 AU,
and any $R_{\rm out}$ larger than 220 AU is acceptable.
Therefore, we fixed $R_{\rm in}$ at 0.1 AU and $R_{\rm out}$ at 310 AU.

The local line width has two components:
turbulent component, $Dv_{\rm t}$, and thermal component.
The turbulent component can be measured in principle
from the spectrum at the edge of the disk,
but our velocity resolution is too coarse.
We assumed $Dv_{\rm t}$ = 0.2 km s$^{-1}$ as a representative value
(Dutrey et al. 2007).
The resulting PV diagram is not very sensitive to $Dv_{\rm t}$
because the line width is unresolved in the outer region
and is dominated by the thermal component in the inner region.
For the thermal component,
the kinetic temperature was assumed to have a power-law profile,
\begin{equation}
T_{\rm K}(r) \propto r^{-q}.
\end{equation}
We assumed $T_{\rm K}$(1 AU) = 500 K,
which is a good representative value for the luminosity of IRAS 4A2.
We also assumed $q$ = 0.6,
which is typical of circumstellar disks of young stellar objects
(Dutrey et al. 2007).

The intensity distribution of the model was convolved with the beam,
and PV diagrams along the disk major axis
were generated for a large number of free-parameter sets.
Best-fit model was searched for using the least-squares fitting
over the data points with intensities higher than 3$\sigma$,
where $\sigma$ is the rms noise level.
The best-fit model (Figures 3(b) and 4) has $v_\phi$(1 AU) = 8.2 km s$^{-1}$
(or $v_\phi$(20 AU) = 1.8 km s$^{-1}$).
Most of the uncertainty in $v_\phi$(1 AU)
comes from the statistical noise in the NH$_3$ data.

To understand how sensitive the velocity is to the assumed parameter values,
we tried models with alternative values.
Varying $Dv_{\rm t}$ by a factor of 2
makes changes in $v_\phi$ by about 1\%.
Varying $T_{\rm K}$ by a factor of 2
makes changes in $v_\phi$ by about 2\%.
Varying $q$ in the range 0.3--0.7 (Dartois et al. 2003)
makes changes in $v_\phi$ by about 2\%.
The uncertainty in the distance, 235 $\pm$ 18 pc (Hirota et al. 2008),
gives an uncertainty in $v_\phi$ to be about 2\%.
The uncertainty in the inclination angle, 10\fdg7 $\pm$ 2\fdg7,
introduces an uncertainty in $v_\phi$ at about 4\% level.
Taking all the above in consideration,
the best-fit model has $v_\phi$(1 AU) = 8.2 $\pm$ 1.3 km s$^{-1}$
and $a$ = 1.94 $\pm$ 0.07.

The distribution of emission in the observed PV diagrams
is somewhat broader than that of the best-fit model PV diagram (Figure 3).
This difference suggests that the power-law disk model
has some oversimplifications.
To assess how good is the velocity profile derived above,
some alternative (if not well justified) models were considered.
For example, we tried models with Gaussian surface-brightness profiles,
which can mimic the effect of NH$_3$ depletion in the inner region,
and the best-fit $v_\phi$(1~AU) is $\sim$7.4 km s$^{-1}$,
well within the uncertainty interval
of the value from the power-law disk model.
We also tried models with Gaussian line-width profiles,
which can mimic the effect of line broadening owing to optical depth effects.
In this case, the fit becomes relatively less sensitive to $v_\phi$,
and the best-fit value is $\sim$9.4 km s$^{-1}$.
Therefore, we conclude that our estimate of $v_\phi$ is robust.
The reason for this robustness is
that, once Keplerian rotation is assumed,
the velocity profile is essentially determined
by the angle of the diagonal high-intensity ridge in the PV diagram.

For the disks of relatively evolved young stellar objects,
the assumption of Keplerian rotation works reasonably well
(Dutrey et al. 2007).
For young protostars like IRAS 4A2,
the validity of this assumption is not guaranteed.
The rotation profile of the disks of actively accreting protostars
can be affected by disk self-gravity, gas pressure gradient,
magnetic forces, and other effects.
However, the deviation from Keplerian rotation
could not be tested with our NH$_3$ data
because the resolution is not good enough.
Future observations with a higher resolution
(both spatial and spectral resolutions)
will be able to provide more accurate information on the rotation profile.

\subsection{Mass and Age}

From the best-fit rotation curve of the IRAS 4A2 disk,
the mass of the central protostar is 0.08 $\pm$ 0.02 $M_\odot$.
As expected for a Class 0 protostar,
this mass is much smaller than the mass of the envelope
($\sim$8 $M_\odot$; Enoch et al. 2009).
Considering that the mass of IRAS 4A2 is already higher
than the minimum mass required for stable nuclear fusion of hydrogen
(0.075 $M_\odot$; Chabrier \& Baraffe 1997),
it will definitely become a hydrogen-burning star.

The bolometric luminosity, $L_{\rm bol}$, of the IRAS 4A binary system
is about four times that of the Sun (Enoch et al. 2009).
Since the binary components have never been resolved in the infrared band,
the luminosity of each protostar is not clear.
While IRAS 4A1 is brighter in dust continuum,
IRAS 4A2 seems to be brighter in molecular lines
and drives a more powerful outflow (Choi 2005; Choi et al. 2007).
To make rough estimates of related physical quantities,
we assume that about half of the luminosity
can be attributed to IRAS 4A2 with a nominal uncertainty of about 50\%,
i.e., $L_{\rm bol}$(IRAS 4A2) $\approx$ 1.9 $\pm$ 0.9 $L_\odot$.
Assuming that the radius of the protostar
is about twice that of the Sun (Stahler 1988),
the rate of mass accretion to the protostar
is about (1.6 $\pm$ 0.9) $\times$ 10$^{-6}$ $M_\odot$ yr$^{-1}$,
which is consistent with what is predicted
by the standard theory of star formation (Shu et al. 1987).

If the accretion has been steady
since the onset of the gravitational collapse,
the age is $\sim$50,000 yr
(with an uncertainty interval of 30,000--90,000 yr).
The mass and age of IRAS 4A2 agree well
with what were expected for Class 0 protostars (Young \& Evans 2005).
This age makes IRAS 4A2 one of the youngest stellar objects
with its mass directly determined
by the kinematics of the circumstellar disk.
Considering that the lifetime of Class 0 sources in Perseus
is $\sim$320,000 yr (Evans et al. 2009),
IRAS 4A2 is extremely young even among the Class 0 protostars.
If we further assume
that IRAS 4A2 continues to accumulate mass at this rate for 540,000 yr,
the typical lifetime of a protostar (Evans et al. 2009),
it will reach a mass of 0.9 $\pm$ 0.5 $M_\odot$.
Therefore, IRAS 4A2 can be considered an analog of the proto-Sun
(but in a binary system, unlike the Sun).

The age (and eventual mass) in the above paragraph
depends on the temporal variation of accretion rate,
which we assumed to be steady.
Since the luminosity and mass accretion rate of IRAS 4A2
are consistent with what are expected from simple theories,
the assumption of steady accretion is reasonable.
However, to explain the existence
of a large number of underluminous protostars,
it has been proposed that the accretion rate may fluctuate significantly
(Kenyon et al. 1990; Dunham et al. 2010).
If IRAS 4A2 has previously experienced an episodic increase of accretion rate
and is currently in a quiescent state,
the age can be much shorter than 50,000 yr.

\subsection{Evolutionary Indicators}

It is interesting to compare the mass derived above
with evolutionary indicators.
However, this comparison is not straightforward for several reasons.
First, the indicators from observations are known
only for the whole IRAS 4A binary system, not for the individual protostars.
Second, at this early stage of evolution,
the binary components are not necessarily coeval.
That is, the evolutionary indicators of the two components,
such as $T_{\rm bol}$, can be different.
Third, model evolutionary tracks are calculated
only for the case of single-star formation,
and these models assume certain values of physical parameters,
such as the initial mass of the dense cloud core
or the final mass of the star,
but these parameters are difficult to determine from observations.

As a simple example, we compare the evolutionary indicators
given by Enoch et al. (2009) and Young \& Evans (2005).
(For a more detailed discussion, see Froebrich 2005.)
For the IRAS 4A binary system, $T_{\rm bol}$ and $L_{\rm bol}$
are 51 $\pm$ 17 K and $\sim$4.2 $L_\odot$, respectively (Enoch et al. 2009).
We assume that IRAS 4A2 has $T_{\rm bol}$ $\approx$ 51 K
and $L_{\rm bol}$ $\approx$ 1.9 $L_\odot$ (see Section 4.2).
Among the evolutionary tracks in Young \& Evans (2005),
we take the one for the initial core mass of 3 $M_\odot$,
which may be relevant to the case of binary formation
in an $\sim$8 $M_\odot$ envelope (Enoch et al. 2009).
From this model, the $T_{\rm bol}$ and $L_{\rm bol}$
of a 0.08 $\pm$ 0.02 $M_\odot$ protostar
are 60 $\pm$ 10 K and 7.7 $\pm$ 2.1 $L_\odot$, respectively.
The temperature agrees within the uncertainty,
while the observed luminosity is somewhat smaller than the model value.
For a better comparison,
it is necessary to measure these indicators for IRAS 4A2
separately from its binary companion
and calculate model values specific to the case of IRAS 4A2.

\acknowledgements

We thank Jeong-Eun Lee for helpful discussions and suggestions.
M.C. was supported by the International Research \& Development Program
of the National Research Foundation of Korea (NRF)
funded by the Ministry of Education, Science and Technology (MEST) of Korea
(grant number: K20901001400-09B1300-03210, FY 2009).
The National Radio Astronomy Observatory is
a facility of the National Science Foundation
operated under cooperative agreement by Associated Universities, Inc.

\end{document}